\documentclass[twocolumn,aps,showpacs,preprintnumbers,amsmath,amssymb]{revtex4}
\usepackage{array,graphicx,color,fancyhdr}
\usepackage{bm}

\begin{document}
\pagestyle{fancy}
\lhead{}
\lfoot{}
\cfoot{ LA-UR 10-02872}
\rhead{ LA-UR 10-02872}
\preprint{W. T. Buttler and S. K. Lamoreaux,  LA-UR 10-02872}
\title{\color{blue} Quantum dense coding without entanglement}

\author{W. T. Buttler,$^1$ S. K. Lamoreaux,$^2$ and J. R. Torgerson$^1$}

\affiliation{$^1$Los Alamos National Laboratory \\
Physics Division (P-23), MS H803 \\
Los Alamos, NM 87545 \\ \\
$^2$Yale University \\
Physics - SPL, PO Box 208120 \\
New Haven, CT 06520-8120}

\begin{abstract}

We describe a quantum cryptography protocol with up to twenty four-dimensional ($\mathcal{D} =4$) states generated by a polarization-, phase- and time-encoding transmitter. This protocol can be experimentally realized with existing technology, drawing from time-encoded and polarization-encoded systems. The protocol is error tolerant and has a quantum bit-rate of 2 per transmission/detection, which when combined with state detection efficiency yields a qubit efficiency of up to 1 or double that of BB84-{\it like} protocols.

\end{abstract}
\date{May 3, 2010}
\maketitle

We investigate the quantum information implications of forming a single-photon superposition of time, phase and polarization states. To our knowledge, the mutual effects of combining multiple basis states of a single photon has has not been fully analyzed, although it can be argued that Bennett established polarization, phase and time encoding in 1992 \cite{b92} when he showed that quantum key distribution (QKD) could be effected through use of a pair of Franson's unbalanced Mach-Zehnder (uMZ) interferometers \cite{franson89, franson91}. While the encoding techniques employed in fiber based BB84 \cite{BB84} require an overlap of the temporal and polarization superposition modes to generate a random key, typically only one polarization mode has been utilized, as required for good interferometric visibility \cite{spatial}. This investigation shows that an inclusion of extended polarization coding, together with the typical fiber QKD elements of phase and time, allows creation of multi-state superpositions of single photons that can be simply exploited to effect efficient and secure QKD---quantum dense coding without entanglement.
\begin{figure}[b] 
\includegraphics[width=1.5 in,height=!,angle=0]{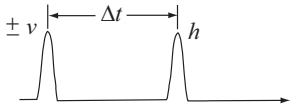}
\caption{Example of a single-photon superposition, a superposition in time, polarization and phase.}
\label{superP}
\end{figure}

The existence of quantum dense coding systems have been discussed in theory \cite{bp01, bp02}, but no practical non-entangled quantum cryptography (QC) concept with {\it Hilbert} dimension $\mathcal{D} > 2$ has been presented or realized. By including polarization mode encoding with the standard time and phase outputs from fiber based BB84, a 4-level single-photon superposition is simply created.

An example of the common fiber based BB84 superposition is shown in Fig. \ref{superP}, and is physically described as
\begin{equation}
   | \psi_0 \rangle = \frac{| h \rangle | t \rangle \pm | v \rangle | t^\prime \rangle}{\sqrt{2}},
\label{psi0}
\end{equation}
where $\psi_0$ is a temporal and polarization superposition of plus-or-minus vertical polarization ($\pm v$) delayed by $\Delta t$ following horizontal polarization ($h$), and where $| t \rangle$ identifies the portion of the superposition at time $t$, and $| t^\prime \rangle \equiv t^\prime = t + \Delta t$. This type of superposition is commonly formed in fiber based BB84 quantum cryptography applications \cite{muller,breguet,townsend1,townsend2}, and the $\pm 1$ (phase of $\phi = 0$ or $\pi$) preceding the $v$ is usually added through phase-modulation (fiber based BB84 usually also randomly adds $\phi = \pi/2$ and $3 \pi/2$ to create uncertainty and enable the quantum-key exchange).

The Eq. \ref{psi0} superpositions can be manipulated to create the four fundamental single-photon basis states that can be expanded into twenty 4-level states in total, with the four primary states presented below:
\begin{eqnarray}
   | \psi_1^{(1)} \rangle & = & | h \rangle | t \rangle,~| \psi_2^{(1)} \rangle = | v \rangle | t \rangle, \nonumber \\
   | \psi_3^{(1)} \rangle & = & | h \rangle | t^\prime \rangle,~| \psi_2^{(1)} \rangle = | v \rangle | t^\prime \rangle,
   \label{d4-1}
\end{eqnarray}
where we use the {\it bra-ket} notation such that $\langle \psi_i^{(1)} | \psi_j^{(1)} \rangle = \langle t | \langle p | q \rangle | t^\prime \rangle = \langle t | \delta_{pq} | t^\prime \rangle = \delta_{pq} \delta_{t t^\prime} = 0$ if either $p \ne q$ or $t \ne t^\prime$, but $\delta_{pq} \delta_{t t^\prime} = 1$ when $p = q$ and $t = t^\prime$, for $p \in \{ h,v \}$ and $q \in \{h,v \}$, as usual.

The mathematics are simplified with definitions
\begin{equation}
   | H \rangle \equiv | h \rangle | t \rangle,~| V \rangle \equiv | v \rangle | t \rangle,~| H^\prime \rangle \equiv | h \rangle | t^\prime \rangle,~| V^\prime \rangle \equiv | v \rangle | t^\prime \rangle. \nonumber
\end{equation}
These primed and unprimed states in linear combinations form left-and-right diagonal, or left-and-right circular, states in the primed or unprimed time coordinates:
\begin{eqnarray}
   D & = & \frac{ H + V }{\sqrt{2}},~D^\prime = \frac{ H^\prime + V^\prime }{\sqrt{2}}, \nonumber \\
   \bar{D} & = & \frac{ H - V }{\sqrt{2}},~\bar{D}^\prime = \frac{ H^\prime - V^\prime }{\sqrt{2}}, \nonumber \\
   R & = & \frac{ H + i V }{\sqrt{2}},~R^\prime = \frac{ H^\prime + i V^\prime }{\sqrt{2}}, \nonumber \\
   L & = & \frac{ H - iV }{\sqrt{2}},~L^\prime = \frac{ H^\prime - i V^\prime }{\sqrt{2}}.
\end{eqnarray}
These definitions, together with the techniques for finding mutually unbiased bases (MUBs) presented in \cite{wooters,klappenecker,brierley}, are used to specify the remaining four $\mathcal{D} =4$ MUBs for this non-entangled single-photon 4-level system:
\begin{eqnarray}
   | \psi_{1}^{(2)} \rangle & = & \frac{ | D \rangle + | D^\prime \rangle }{\sqrt{2}},~| \psi_{2}^{(2)} \rangle = \frac{ | D \rangle - | D^\prime \rangle }{\sqrt{2}}, \nonumber \\
   | \psi_{3}^{(2)} \rangle & = & \frac{ | \bar{D} \rangle - | \bar{D}^\prime \rangle }{\sqrt{2}},~| \psi_{4}^{(2)} \rangle = \frac{ | \bar{D} \rangle + | \bar{D}^\prime \rangle }{\sqrt{2}}, \label{d4-2}
\end{eqnarray}
\begin{eqnarray}
   | \psi_{1}^{(3)} \rangle & = & \frac{ | D \rangle + i | \bar{D}^\prime \rangle }{\sqrt{2}},~| \psi_{2}^{(3)} \rangle = \frac{ | D \rangle - i | \bar{D}^\prime \rangle }{\sqrt{2}}, \nonumber \\
   | \psi_{3}^{(3)} \rangle & = &  \frac{ | \bar{D} \rangle - i | D^\prime \rangle }{\sqrt{2}},~| \psi_{4}^{(3)} \rangle = \frac{ | \bar{D} \rangle + i | D^\prime \rangle }{\sqrt{2}},
   \label{d4-3}
\end{eqnarray}
\begin{eqnarray}
   | \psi_{1}^{(4)} \rangle & = & \frac{ | R \rangle - | L^\prime \rangle }{\sqrt{2}},~| \psi_{2}^{(4)} \rangle = \frac{ | L \rangle - | R^\prime \rangle }{\sqrt{2}}, \nonumber \\
   | \psi_{3}^{(4)} \rangle & = & \frac{ | R \rangle + | L^\prime \rangle }{\sqrt{2}},~| \psi_{4}^{(4)} \rangle = \frac{ | L \rangle + | R^\prime \rangle }{\sqrt{2}},
   \label{d4-4}
\end{eqnarray}
\begin{eqnarray}
   | \psi_{1}^{(5)} \rangle & = & \frac{ | R \rangle + i | R^\prime \rangle }{\sqrt{2}},~| \psi_{2}^{(5)} \rangle = \frac{ | R \rangle - i | R^\prime \rangle }{\sqrt{2}}, \nonumber \\
   | \psi_{3}^{(5)} \rangle & = & \frac{ | L \rangle - i | L^\prime \rangle }{\sqrt{2}},~| \psi_{4}^{(5)} \rangle = \frac{ | L \rangle + i | L^\prime \rangle }{\sqrt{2}}.
   \label{d4-5}
\end{eqnarray}
These single-photon states form twenty 4-level {\it qudits}, in contrast with 2-level {\it qubits}, and it is evident that $ | \langle \psi_i^m | \psi_j^n \rangle |^2 = \delta_{ij}$ for all $\{i,j\}$ when $m = n$, and it is easily demonstrated that $ | \langle \psi_i^m | \psi_j^n \rangle |^2 = 1/4$ for all $\{i,j\}$ when $m \ne n$, proving that the $\psi_i^m$ form five 4-level single-photon MUBs. Other non-optimal basis sets can be found, but these five bases represent the maximum number of 4-level bases that exist when $\mathcal{D}$ is a power of a prime \cite{brierley}.

An experimental realization of these single-photon states could be used to form an efficient and secure QC protocol if the states can be optimally detected. For example, 4-level QC systems can tolerate error rates on the qudits of 0.25/qudit for individual attacks \cite{fuchs}, and error rates of 0.1893/qudit for coherent attacks \cite{Lo2,shor,tittel}, e.g., as demonstrated in \cite{bruss} and especially Table 1 in \cite{cerf}. These tolerable qudit error rates are to be compared with the maximum allowable error rates for single-photon 2-level QC systems of 0.1464/qubit for individual attacks and 0.1100/qubit for coherent attacks (the 0.1100/qubit has been shown to be the maximum allowable error rate for {\it unconditional} qubit security for BB84 \cite{shor}).

\begin{figure}[b] 
\includegraphics[width=3.0 in,height=!,angle=0]{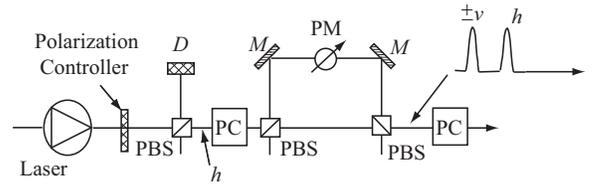}
\caption{Optical model with degrees of freedom needed to create 4-level single photons. Optics include a laser source that transmits both a bright timing-pulse and single photons onto the quantum channel, a polarization controller, three polarizing beamsplitters (PBSs), a phase modulator (PM), and two Pockels cells (PCs). The early position exits the uMZ with $h$-polarization, and the late position with $\pm v$-polarization.}
\label{transmitter}
\end{figure}

One technique to prepare the $\mathcal{D} =4$ states is modeled with the Fig. \ref{transmitter} optics. In this approach a $h$ photon is transmitted beyond the first polarizing beamsplitter (PBS) and encounters the first Pockels cell (PC) that can be modulated to transmit either a horizontal, vertical, or positive diagonal polarized photon ($h$, $v$ or $d = (h + v)/\sqrt{2}$) as the input state to the uMZ \cite{left-d}. For the case that the first PC rotates the $h$ polarization to $d$, the second PBS causes an $h$ and $v$ superposition of the single photon to form, where the $v$ position travels the upper path and the $h$ position the lower path through the uMZ. For the case that the first PC transmits $h$ or $v$, the $h$ polarization will pass directly through the PBS to the lower-arm of the uMZ, and the $v$ polarization will be reflected by the PBS to the upper-arm of the uMZ. For both cases, the lower-arm includes no optics to alter the state of the photon (or photon position) traveling this path, but the upper-arm contains a phase modulator (PM) that randomly adds a phase delay of $\phi = 0$ or $\pi$ to the upper-arm photon (or photon position). The uMZ is complete with a third PBS that transmits $h$ and reflects $v$ so that for the first case the Eq. \ref{psi0} state $\psi_0$ is transmitted onto the quantum channel, i.e., the transmitted state is $\pm v$ delayed by $\Delta t$ following $h$, where $t$ and $t^\prime = t + \Delta t$ are as defined earlier, and $\Delta t$ is the time difference between the long and short arms of the uMZ \cite{delta-t}, but for the second case either $h$ is transmitted at time $t$, or else $v$ is transmitted at time $t^\prime$. For both cases the second PC is modulated to rotate the $t$ or $t^\prime$ polarization to $h,~v,~d,~\bar{d},~r$, or $l$ ($r$ and $l$ are left and right circular polarization: $r = (h + iv)/\sqrt{2}$ and $l = (h - iv)/\sqrt{2}$), where we suppose that the additional phase of $\pm i$ can be added as needed at the $t^\prime$ states through quarter-wave ($Q$) modulation of the first PC.

\begin{figure}[b] 
\includegraphics[width=3.25 in,height=!,angle=0]{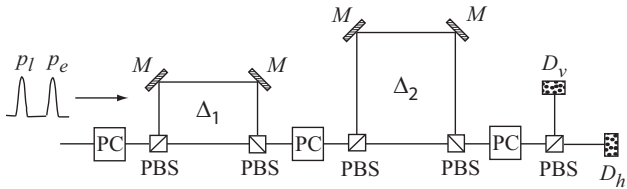}
\caption{An optical model that permits discrimination of the $\mathcal{D} =4$ states presented in Eqs. \ref{d4-1} and \ref{d4-2} to \ref{d4-5}. Optics include three PCs that modulate the early ($p_e$) and late ($p_l$) superpositions, two time-delay operators composed of four PBSs and four {\it M}s that temporally delay $v$ positions relative to $h$ positions ($\widehat{\Delta}_1 \equiv t^\prime - t = \Delta t$ and $\widehat{\Delta}_2 \equiv 2 \Delta t$), and a PBS that follows the third PC and projects $h$ and $v$ onto $D_h$ and $D_v$.} \label{d4detection}
\end{figure}

The Fig. \ref{transmitter} optics, with its degrees of freedom defined by the two PCs (symmetry operators), can be used to create any of the twenty $\psi_i^m$ states. Based on, (1) security of 4-level systems to individual (25\% to 26.66\%) and coherent attacks (18.93\%), (2) the marginal increase in security against individual attacks for twenty 4-level states versus eight 4-level states (1.66\%) relative to the dense-coding bit rate efficiencies [$\rho(\text{2-bases}) = 1,~\rho(\text{3-bases}) = 2/3,~\rho(\text{4-bases}) = 1/2,~\rho(\text{5-bases}) = 2/5$], and (3) because of the Shannon reconciliation limit based on initial error rates, we only consider eight 4-level states in two bases: bases $\psi_i^{m}$ for $i \in \{1,2,3,4\}$, and $m \in \{2,4\}$ (Eqs. \ref{d4-2} and \ref{d4-4}). These eight states ($\psi_i^2$, and $\psi_i^4$) can be resolved in point-to-point QKD with interferometric techniques and the BLT02-{\it like}~\cite{blt02} optics seen in Fig. \ref{d4detection}, with a protocol detection efficiency of $\eta = 1/2$ if only two of the basis states are used \cite{other-det}.

The efficient detection of the $\psi_i^{(2 ^\lor 4)}$ states can be accomplished by modeling the three PCs in Fig. \ref{d4detection} as rotation operators $\widehat{P}_{1}^{(m)}$, $\widehat{P}_{2}^{(m)}$, and $\widehat{P}_{3}^{(m)}$, respectively, where the $m$ superscript denotes either the $\psi^2$ or $\psi^4$ basis, and by letting the two delay legs $\Delta_1$ and $\Delta_2$ be denoted by operators $\widehat{\Delta}_1$ and $\widehat{\Delta}_2$ \cite{delay-legs}. In this model the early ($p_e$) and late ($p_l$) positions of the $\mathcal{D} =4$ states enter the optical system and encounter the operators in order from left to right: $\widehat{P}_{1}^{(m)}$, $\widehat{\Delta}_1$, $\widehat{P}_{2}^{(m)}$, $\widehat{\Delta}_2$, to the final $\widehat{P}_{3}^{(m)}$ rotation operator, and then arrive at the PBS that projects $v$ polarizations onto detector $D_v$, and $h$ polarizations onto detector $D_h$. With this optical model, resolution of these states requires either half-wave operations of $\widehat{P}_{1}^{(2)} \equiv \widehat{H}^{(45)}(t) + \widehat{H}^{(-45)}(t + \Delta t)$ for the $p_e$ and $p_l$ polarization positions (times $t$ and $t^\prime$), or quarter-wave operations of $\widehat{P}_{1}^{(4)} \equiv \widehat{Q}$, at both $t$ and $t^\prime$ for the $p_e$ and $p_l$ polarization positions. Except for these operational differences, the remaining operator elements to resolve the $\psi_i^{2}$ and $\psi_i^4$ states are fixed (thus we drop the superscripts), with $\widehat{P}_{2} \equiv  \widehat{H}^{(90)}(t + 2 \Delta t) +  \widehat{H}^{(45)}(t + \Delta t) +  \widehat{H}^{(90)}(t)$ rotation operations, and $\widehat{P}_{3} \equiv \widehat{I}(t + \Delta t) + \widehat{H}^{(45)}(t + 2 \Delta t) + \widehat{I}(t + 3 \Delta t)$.

Using the optical models defined by Figs. \ref{transmitter} and \ref{d4detection}, and operator notation as described, we find ordered operations that allow resolution of the Eqs. \ref{d4-2} and \ref{d4-4} states
\begin{eqnarray}
   %
   \widehat{O} \big [ \psi^{(2)} \big ] & \equiv & \widehat{P}_{3} \widehat{\Delta}_2 \widehat{P}_{2} \widehat{\Delta}_1  \widehat{P}_{1}^{(2)}, \nonumber \\
   \widehat{O} \big [ \psi^{(4)} \big ] & \equiv & \widehat{P}_{3} \widehat{\Delta}_2 \widehat{P}_{2} \widehat{\Delta}_1  \widehat{P}_{1}^{(4)}.
   \label{fD4Ops}
\end{eqnarray}
The $\widehat{O}\big [ \psi^{(2)} \big ]$ and $\widehat{O} \big [ \psi^{(4)} \big ]$ operate on the arriving states to direct them in a self consistent manner to the appropriate time slots and detectors to resolve their respective bases. For example,
\begin{eqnarray}
   \widehat{O} \big [ \psi^{(2)} \big ] | \psi_1^{(2)} \rangle & \Leftrightarrow & \widehat{O} \big [ \psi^{(4)} \big ] | \psi_1^{(4)} \rangle \rightarrow | v \rangle | t + 2 \Delta t \rangle \equiv V^{\prime \prime} \nonumber \\
   & \Rightarrow & \psi_1^2 \land \psi_1^4 \mapsto D_v( t + 2 \Delta t), \nonumber \\
   \widehat{O} \big [ \psi^{(2)} \big ] | \psi_2^{(2)} \rangle & \Leftrightarrow & \widehat{O} \big [ \psi^{(4)} \big ] | \psi_2^{(4)} \rangle \rightarrow | h \rangle | t + 2 \Delta t \rangle \equiv H^{\prime \prime} \nonumber \\
   & \Rightarrow & \psi_2^2 \land \psi_2^4 \mapsto D_h( t + \Delta t), \nonumber \\
   \widehat{O} \big [ \psi^{(2)} \big ] | \psi_3^{(2)} \rangle & \Leftrightarrow & \widehat{O} \big [ \psi^{(4)} \big ] | \psi_3^{(4)} \rangle \rightarrow | v \rangle | t + 3 \Delta t \rangle \equiv V^{\prime \prime \prime} \nonumber \\
   & \Rightarrow & \psi_3^2 \land \psi_3^4 \mapsto D_v( t + 3 \Delta t), \nonumber \\
   \widehat{O} \big [ \psi^{(2)} \big ] | \psi_4^{(2)} \rangle & \Leftrightarrow & \widehat{O} \big [ \psi^{(4)} \big ] | \psi_4^{(4)} \rangle \rightarrow | h \rangle | t + \Delta t \rangle \equiv H^\prime \nonumber \\
   & \Rightarrow & \psi_4^2 \land \psi_4^4 \mapsto D_h( t + \Delta t),
   \label{operators}
\end{eqnarray}
confirming the efficient detection scheme. In contrast, a measurement any of these eight arriving states in the other basis causes random detections in consistent statistical distributions at the correct times and detectors. This realization is that
\begin{equation}
   \widehat{O} \big [ \psi^{(m)} \big ] | \psi_i^{(n)} \rangle = \frac{ a | H^\prime \rangle + \sqrt{2} b | D^{\prime \prime} \rangle + c | V^{\prime \prime \prime} \rangle }{2},
\end{equation}
where the single prime coordinate implies detection at time $t^\prime = t + \Delta t$, the double prime coordinate implies detection at time $t^{\prime \prime} = t + 2 \Delta t$, and the triple prime coordinate implies detection at time $t^{\prime \prime \prime} = t + 3 \Delta t$, when $m \ne n$, and where $| a | = | b | = | c | = 1$. From this result it is clear that a measurement of any of these eight states in the incorrect basis will result in random detections on either $D_h$ at time $t^\prime$, $D_h$ or $D_v$ at $t^{\prime \prime}$, or on $D_v$ at time $t + 3 \Delta t$, each with equal detection probability $1/4$ for each detector at each time.

With these physics models, a quantum dense key exchange can be accomplished. The exchange of a secret with the $\psi^2$ and $\psi^4$ basis sets (eight states, two bases) proceeds similarly to BB84. That is, Alice could randomly prepare single photons in one of the eight $\psi_i^2$ or $\psi_i^4$ states and direct them along the quantum channel to Bob who randomly measures the arriving states in one of the two bases as they arrive ($\widehat{O} [ \psi^2 ]$ or $\widehat{O} [ \psi^4 ]$ operational measurement sets). Bob then publicly announces his basis choice and the relative times $t$ of his detections, but not the absolute times of his detections that includes $t$ and some multiple of $\Delta t$. Alice then publicly announces which of his measurements correlated to the transmission basis, and Bob discards the half of his measurements within the wrong basis to complete the sifting process. Alice and Bob then reconcile \cite{cascade, winnow, winnow2, lodewyck} and privacy amplify \cite{winnow,privamp} their key to complete the exchange.

A proof of unconditional security is beyond the scope of this effort \cite{scarani}. However, some clarifying observations can be made. For example, this 4-level entanglement-free concept is to 4-level entangled systems as BB84 is to E91. Therefore, any unconditional security proof of 4-level entangled systems will apply to this proposition. It is therefore observed that the $d$-level security proofs for individual (25.00\% error rate for two 4-level bases) and coherent attacks (18.93\%) do apply. It is also noted that the coherent limit of 11.00\% is the unconditional limit calculated for BB84/E91. Thus it is likely that 18.93\% is the unconditional security limit for 4-level systems, or a close approximation.

It is also noted that these concepts can be implemented with true single-photon sources, or with weak coherent pulses (WCP). The WCP are easy to create, but have been shown to limit range of operation as compared with true single photon QC systems. Regarding WCP, this system allows monitoring of photon number in the error rate as two detectors are used, and detections are observed on the two detectors at three unique times on the $D_h$ and $D_v$ detectors. Thus multiple detections on $D_h$ or $D_v$ at $t^\prime$, $t^{\prime \prime}$ or $t^{\prime \prime \prime}$, or detections on $D_v$ at $t^\prime$, or on $D_h$ at $t^{\prime \prime \prime}$, give clues about fidelity and possible attacks. It is also clear that these concepts can be implemented with decoy states \cite{hwang} which provides a test of the presence of certain attacks on the quantum key.

Like Cerf, et al. \cite{cerf}, we conclude that a two-bases protocol is preferred to a $\mathcal{D} + 1$ basis protocol because its maximum error rate is only slightly lower, and because each detection leads to two bits, which when combined with the basis choice selection efficiency of $\eta = 1/2$, implies a rate of 1-bit per transmission with ideal single-photon QC in a lossless channel (which of course does not exist): the secret key rate is larger than for $\mathcal{D} = 2$ BB84, or for multiple bases in $\mathcal{D} = 4$ QC. Further, the amount of key compression in the privacy amplification is lower due to the higher tolerance of errors. Cerf, et al., also showed that the tolerable error approaches $1/2$ when $\mathcal{D} \rightarrow \infty$, and we have shown in a general way how to increase the dimension with polarization, phase and time encoding of larger and larger plane-wave dimensions.

We have described a new high-dimensional quantum key distribution system that employs the simultaneous use of several single photon basis states. These states include the polarization states as used in, for example, BB84, combined with time-domain encoding, as used with Franson's uMZs, and phase encoding. Extension of BB84-like protocols to include an infinitely large plane wave basis states allows the generation of high-dimensional quantum states that can be realized with existing quantum cryptography experimental apparatus.

\begin{center} {\bf {APPENDIX}} \end{center}

Other 4-level bases can be found and used to effect a quantum key exchange. For example, the Eq. \ref{psi0} superpositions can be manipulated to create three sets of mutually unbiased 4-level single-photon basis states---twelve states in total---through symmetry operations (rotations) of the early or late polarizations, as demonstrated in Eqs. \ref{d42-1} to \ref{d42-3}:
\begin{eqnarray}
   | \chi_{1}^{(1)} \rangle & = & \frac{| H \rangle + | V^\prime \rangle}{\sqrt{2}} \text{,~} | \chi_{2}^{(1)} \rangle = \frac{| H \rangle - | V^\prime \rangle}{\sqrt{2}}, \nonumber \\
   | \chi_{3}^{(1)} \rangle & = & \frac{| V \rangle + | H^\prime \rangle}{\sqrt{2}} \text{,~} | \chi_{4}^{(1)} \rangle = \frac{| V \rangle - | H^\prime \rangle}{\sqrt{2}}, \label{d42-1}
\end{eqnarray}
\begin{eqnarray}
   | \chi_{1}^{(2)} \rangle & = & \frac{| H \rangle + | H^\prime \rangle}{\sqrt{2}} \text{,~} | \chi_{2}^{(2)} \rangle = \frac{| H \rangle - | H^\prime \rangle}{\sqrt{2}}, \nonumber \\
   | \chi_{3}^{(2)} \rangle & = & \frac{| V \rangle - | V^\prime \rangle}{\sqrt{2}} \text{,~} | \chi_{4}^{(2)} \rangle = \frac{| V \rangle + | V^\prime \rangle}{\sqrt{2}}, \label{d42-2}
\end{eqnarray}
\begin{eqnarray}
   | \chi_1^{(3)} \rangle & = & | D \rangle \text{,~~~~} \, \, | \chi_2^{(3)} \rangle = \bar{D} \rangle, \nonumber \\
   | \chi_3^{(3)} \rangle & = & | D^\prime \rangle \text{,~~~~} | \chi_4^{(3)} \rangle = \bar{D}^\prime \rangle.
   \label{d42-3}
\end{eqnarray}
It is evident that $ | \langle \chi_i^m | \chi_j^n \rangle |^2 = \delta_{ij}$ for all $\{i,j\}$ when $m = n$, but that $ | \langle \chi_i^m | \chi_j^n \rangle |^2 = 1/4$ for all $\{i,j\}$ when $m \ne n$, proving that the $\chi_i^m$ form three-sets of mutually unbiased 4-level single-photon basis states for $m,n \in \{1,2,3\}$.

To prove that the states in Eqs. \ref{d42-1} to \ref{d42-3} represent a complete set of dimension $\mathcal{D} = 4$, MUBs, we begin by identifying one pair of maximally non-orthogonal states. For our purposes, we choose the states $| \chi^1 \rangle$ and $| \chi^2 \rangle$ of Eqs. \ref{d42-1} and \ref{d42-2}, although any pair can work equally well. States $\chi^1_i$ and $\chi^2_j$ are also quite general and simple rotations yield other equivalent superpositions, of which a specific example is states $| \chi^1_i \rangle$ and $| \chi^2_j \rangle$ where $| h \rangle \rightarrow | r \rangle$ and $| v \rangle \rightarrow - | l \rangle$.

In any case, what is sought are other solutions of the form
\begin{equation}
	|\lambda_i\rangle = a_i | H \rangle + b_i | V
                      \rangle + c_i | H^\prime \rangle + d_i | V^\prime \rangle,
\label{d4-gen}
\end{equation}
where $a_i = A_i\cdot e^{i\cdot\theta_{ai}} $ and $b_i = B_i \cdot e^{i \cdot \theta_{bi}}$, etc., with the restriction that $| \langle \lambda_i | \lambda_j \rangle |^2 = \delta_{ij}$. As before, the requirement on new states $| \lambda \rangle$ are that they are unbiased with both $| \chi^1_i \rangle$ and $| \chi^2_j \rangle$ $\mathcal{D} = 4$ states, or specifically $| \langle \chi_i^1 | \lambda_j \rangle |^2 = 1/4$ for all $\{i,j\}$ and $| \langle \chi_i^2 | \lambda_j \rangle |^2 = 1/4$ for all $\{i,j\}$. The conditions can be applied to yield
\begin{eqnarray}
   A D \cos(\theta_{ad}) & = & 0, \nonumber \\
   A C \cos(\theta_{ac}) & = & 0, \nonumber \\
   B D \cos(\theta_{bd}) & = & 0, \nonumber \\
   B C \cos(\theta_{bc}) & = & 0,
\end{eqnarray}
where $\theta_{xy} = \theta_x - \theta_y$, and also
\begin{equation}
   A D = B C = A C = B D = 0.
\end{equation}
for any particular $| \lambda \rangle$. (The $i$'s were omitted for clarity.)

This results in the following:
\begin{eqnarray}
   {\rm either\hspace{20pt}} \{A,B\} & = & 0,\\
   {\rm or\hspace{36pt}} \{C,D\} & = & 0,
\end{eqnarray}
and the values for the phase angles $\{ \theta_a, \theta_b, \theta_c, \theta_d \}$ can be decided as a matter of convenience. Thus, the only possible solution for the $\lambda_i$ are the ones labeled $| \chi_i^3 \rangle$ in Eq. \ref{d42-3} (or simple rotations of them).

The efficient detection of the $\chi^m_i$ states proceeds similarly as before.
$\widehat{P}_{1}^{(1)} \equiv \widehat{I}$ at both $t$ and $t^\prime$, but $\widehat{P}_{1}^{(2)} \equiv \widehat{I}(t) + \widehat{H}^{(90)}(t + \Delta t)$ for the $p_e$ and $p_l$ polarization positions (times $t$ and $t^\prime$). The remaining operator elements to resolve the $\chi_i^1$ and $\chi_j^2$ states are fixed, with $\widehat{P}_{2}^{(1)} \Leftrightarrow \widehat{P}_{2}^{(2)} \equiv  \widehat{H}^{(90)}(t + 2 \Delta t) + \widehat{H}^{(45)}(t + \Delta t) +  \widehat{H}^{(90)}(t)$ rotation operations, and $\widehat{P}_{3}^{(1)} \Leftrightarrow \widehat{P}_3^{(2)} \equiv \widehat{I}(t + \Delta t) + \widehat{H}^{(45)}(t + 2 \Delta t) + \widehat{I}(t + 3 \Delta t)$. The $\chi_i^3$ states are resolved when $\widehat{P}_{1}^{(3)} \equiv \widehat{H}^{(45)}$, $\widehat{P}_{2}^{(3)} \equiv \widehat{H}^{(90)}$, and when $\widehat{P}_{3}^{(3)} \equiv \widehat{I}$ for all times.

Using the optical models defined by Figs. \ref{transmitter} and \ref{d4detection}, and operator notation as described, we find ordered operational sets that allow resolution of the Eqs. \ref{d42-1} to \ref{d42-3} states
\begin{eqnarray}
   \widehat{O} \big [ \chi^{(1)} \big ] & \equiv & \widehat{P}_{3}^{(1)} \widehat{\Delta}_2 \widehat{P}_{2}^{(1)} \widehat{\Delta}_1 \widehat{P}_{1}^{(1)}, \nonumber \\
   \widehat{O} \big [ \chi^{(2)} \big ] & \equiv & \widehat{P}_{3}^{(2)} \widehat{\Delta}_2 \widehat{P}_{2}^{(2)} \widehat{\Delta}_1  \widehat{P}_{1}^{(2)}, \nonumber \\
   \widehat{O} \big [ \chi^{(3)} \big ] & \equiv & \widehat{P}_{3}^{(3)} \widehat{\Delta}_2 \widehat{P}_{2^{(3)}} \widehat{\Delta}_1  \widehat{P}_{1}^{(3)}.
   \label{fD42Ops}
\end{eqnarray}
These operators direct the arriving states in a self consistent manner to the appropriate time slots and detectors to resolve their respective bases:
\begin{eqnarray}
   \widehat{O} \big [ \chi^{(1)} \big ] | \chi_1^{(1)} \rangle & \equiv & \widehat{O} \big [ \chi^{(2)} \big ] | \chi_1^{(2)} \rangle \rightarrow -| h \rangle | t + 2 \Delta t \rangle \nonumber \\
   & \Rightarrow & \chi_1^{(1)} \land \chi_1^{(2)} \mapsto D_h( t + 2 \Delta t), \nonumber \\
   \widehat{O} \big [ \chi^{(1)}] | \chi_2^{(1)} \rangle & \equiv & \widehat{O} \big [\chi^{(2)} \big ] | \chi_2^{(2)} \rangle \rightarrow | v \rangle | t + 2 \Delta t \rangle \nonumber \\
   & \Rightarrow & \chi_2^{(1)} \land \chi_2^{(2)} \mapsto D_v( t + 2 \Delta t), \nonumber \\
   \widehat{O} \big [ \chi^{(1)}] | \chi_3^{(1)} \rangle & \equiv & \widehat{O} \big [\chi^{(2)} \big ] | \chi_3^{(2)} \rangle \rightarrow | v \rangle | t + 3 \Delta t \rangle \nonumber \\
   & \Rightarrow & \chi_3^{(1)} \land \chi_3^{(2)} \mapsto D_v( t + 3 \Delta t), \nonumber
\end{eqnarray}
\begin{eqnarray}
   \widehat{O} \big [ \chi^{(1)}] | \chi_4^{(1)} \rangle & \equiv & \widehat{O} \big [\chi^{(2)} \big ] | \chi_4^{(2)} \rangle \rightarrow | h \rangle | t + \Delta t \rangle \nonumber \\
   & \Rightarrow & \chi_4^{(1)} \land \chi_4^{(2)} \mapsto D_h( t + \Delta t),
   \label{operators2}
\end{eqnarray}
and that
\begin{eqnarray}
   \widehat{O}[\chi^{(3)}] | \chi_1^{(3)} \rangle & \rightarrow & H^\prime \Rightarrow \chi_1^{(3)} \mapsto D_h( t + \Delta t), \nonumber \\
   \widehat{O}[\chi^{(3)}] | \chi_2^{(3)} \rangle & \rightarrow & V^{\prime \prime} \Rightarrow \chi_2^{(3)} \mapsto D_v( t + 2 \Delta t), \nonumber \\
   \widehat{O}[\chi^{(3)}] | \chi_3^{(3)} \rangle & \rightarrow & H^{\prime \prime} \Rightarrow \chi_{3}^{(3)} \mapsto D_h( t + 2 \Delta t), \nonumber \\
   \widehat{O}[\chi^{(3)}] | \chi_4^{(3)} \rangle & \rightarrow & V^{\prime \prime \prime} \Rightarrow \chi_4^{(3)} \mapsto D_v( t + 3 \Delta t),
   \label{alpha-operators}
\end{eqnarray}
Measurement of the arriving states in the wrong basis causes random detections in consistent statistical distributions at the correct times and detectors:
\begin{equation}
   \widehat{O} \big [ \chi^{(m)} \big ] | \chi_i^{(n)} \rangle \approx \frac{ | a H^\prime \rangle + \sqrt{2} b | D^{\prime \prime} \rangle + c | V^{\prime \prime \prime} \rangle }{2},
\end{equation}
when $m \ne n$, and where $| a | = | b | = | c | = 1$. This result demonstrates that a measurement of any state in the incorrect basis will result in random detections on either $D_h$ at time $t + \Delta t$, $D_h$ or $D_v$ at $t + 2 \Delta t$, or on $D_v$ at time $t + 3 \Delta t$, each with equal detection probability $p = 1/4$ for each detector at each time. It is also clear that none of the $\chi_i^m$ states are included in the $\psi_i^n$ states. Here we find only three MUBs, but the maximum number of MUBs are the five presented earlier.

\begin{center} {\bf Acknowledgements} \end{center}

The authors thank Steve Brierley, of the University of York, England, for helpful comments on the construction of mutually unbiased bases.


\clearpage


\begin{thebibliography}{100}

\bibitem{b92} C. H. Bennett, Phys. Rev. Lett. {\bf 68}, 3121-3124 (1992).

\bibitem{franson89} J. D. Franson, Phys. Rev. Lett. {\bf 62}, 2205-2208 (1989).

\bibitem{franson91} J. D. Franson, Phys. Rev. A {\bf 44}, 4552-4555 (1991).

\bibitem{BB84} C. H. Bennett and G. Brassard, in {\it Proceedings of the IEEE International Conference on Computers, Signals, and Signal Processing, Bangalore, India, 1984} (IEEE, New York, 1984), pp. 175-179.

\bibitem{spatial} The spatial modes must also overlap to achieve an interferometric quantum exchange. This mode can also be exploited similarly as the polarization modes.

\bibitem{bp01} H. Bechmann-Pasquinucci and W. Tittel, Phys. Rev. A {\bf 61}, 062308 (2000).

\bibitem{bp02} H. Bechmann-Pasquinucci and A. Peres, Phys. Rev. Lett. {\bf 85}, 3313-3316 (2000).

\bibitem{muller} A. Muller, J. Br$\acute{{\text e}}$guet and N. Gisin, Europhys. Phys. Lett. {\bf 23}, 383-388 (1993).

\bibitem{breguet} J. Br$\acute{{\text e}}$guet, A. Muller and N. Gisin, J. Mod. Opt. {\bf 41}, 2405-2412 (1994).

\bibitem{townsend2} P. D. Townsend, J. G. Rarity and P. R. Tapster, Electron. Lett. {\bf 29}, 634-635 (1993).

\bibitem{townsend1} P. D. Townsend J. G. Rarity, and P. R. Tapster, Electron. Lett. {\bf 29}, 1291-1293 (1993).

\bibitem{wooters} W. K. Wooters and B. D. Fields, Annals Phys. {\bf 191}, 363-381 (1989).

\bibitem{klappenecker} A. Klappenecker and M. R\"{o}tteler, Lect. Notes Comput. Sc. {\bf 2984}, 137-144 (2004).

\bibitem{brierley} S. Brierley, S. Weigert, and I. Bengtsson, arXiv:0907.4097 (2009).

\bibitem{fuchs} C. A. Fuchs, N. Gisin, R. B. Griffiths, C. S. Niu, and A. Peres, Phys. Rev. A {\bf 56}, 1163-1172 (1997).

\bibitem{Lo2} H.-K. Lo and H. F. Chau, Science {\bf 283}, 2050-2056 (1999).

\bibitem{shor} P. W. Shor and J. Preskill, Phys. Rev. Lett. {\bf 85}, 441-444 (2000).

\bibitem{tittel} H. Bechmann-Pasquinucci and W. Tittel, Phys. Rev. A {\bf 61}, 062308 (2000).

\bibitem{bruss} D. Bru\ss ~and C. Macchiavello, Phys. Rev. Lett. {\bf 88}, 127901 (2002).

\bibitem{cerf} N. J. Cerf, M. Bourennane, A. Karlsson and N. Gisin, Phys. Rev. Lett. {\bf 88}, 127902 (2002).

\bibitem{left-d} For completeness a left diagonal polarized photon is $\bar{d} = (-h + v)/\sqrt{2}$.

\bibitem{delta-t} The time difference $\Delta t$ is chosen to be long enough to allow the second PC to operate on (modulate the polarizations of) both temporal positions of $\psi_0$, and timing agreement can be accomplished by transmission of an initial bright pulse to announce the beginning of the quantum transmission.

\bibitem{blt02} W. T. Buttler, J. R. Torgerson and S. K. Lamoreaux, Phys. Lett. A {\bf 299}, 38-42 (2002).

\bibitem{other-det} Other efficient detection schemes can be developed. For example, without loss of generality, the $\widehat{\Delta}_2$ can be swapped with $\widehat{\Delta}_1$.

\bibitem{delay-legs} The delay legs are formed with 4 PBSs and 4 $M$s, and are precisely balanced, with respect to the transmission delay leg, so that after operation on the modulated polarizations $\widehat{\Delta}_1$ and $\widehat{\Delta}_2$ delays $v$ polarizations by either $\Delta t$ or $2 \Delta t$ relative to the $h$ polarizations.

\bibitem{cascade} G. Brassard and L. Salvail, Lect. Notes Comput. Sc. {\bf 765}, 410-423 (1994).

\bibitem{winnow} W. T. Buttler, et al., Phys. Rev. A {\bf 67}, 052303 (2003).

\bibitem{winnow2} J. Han and X. Quian, Quant. Inf. Computation {\bf 9}, 0693-0700 (2009).


\bibitem{lodewyck} J. Lodewyck, et al., Phys. Rev. A {\bf 76}, 042305 (2007).

\bibitem{privamp} C. H. Bennett, et al., IEEE Trans. Inf. Theory {\bf 41}, 1915-1923 (1995).

\bibitem{scarani} V. Scarani, et al., Rev. Mod. Phys. {\bf 81}, 1301-1350 (2009).

\bibitem{hwang} W.-Y. Hwang, Phys. Rev. lett. {\bf 91}, 057901 (2003).

\end{thebibliography}
\end{document}